\documentclass[%
 reprint,
 amsmath,amssymb,
 aps,
]{revtex4-1}

\usepackage{graphicx} % Include figure files
\usepackage{dcolumn} % Align table columns on decimal point
\usepackage{bm} % bold math
\usepackage{color}
\usepackage{soul}

\begin{document}

\preprint{APS/123-QED}

\title{Shape Control for Experimental Continuation}

\author{Robin M. Neville}
\author{Rainer M. J. Groh}
\author{Alberto Pirrera}
\email{alberto.pirrera@bristol.ac.uk}
\author{Mark Schenk}
\affiliation{Bristol Composites Institute (ACCIS), Department of Aerospace Engineering, University of Bristol, Bristol, BS8 1TR, United Kingdom}

\begin{abstract}
An experimental method has been developed to locate unstable equilibria of nonlinear structures quasi-statically. The technique involves loading a structure by application of either a force or a displacement at a main actuation point, while simultaneously controlling the overall shape using additional probe points. The method is applied to a shallow arch, and unstable segments of its equilibrium path are identified experimentally for the first time. Shape control is a fundamental building block for the experimental---as opposed to numerical---continuation of nonlinear structures, which will significantly expand our ability to measure their mechanical response.
\end{abstract}

\maketitle

\section{Background}
The force-displacement response of nonlinear structures can be complex and even chaotic. Limit and branch points can partition equilibrium manifolds into stable and unstable segments. In particular, displacement and force limit points change the stability of a structure and inject unstable eigenmodes into the deformation shape, thereby rendering certain segments of the ensuing force-displacement manifolds inaccessible experimentally. This kind of behaviour is observed even in simple structures, such as the shallow arch studied in this paper.

\begin{figure}
  	\centering
     \includegraphics{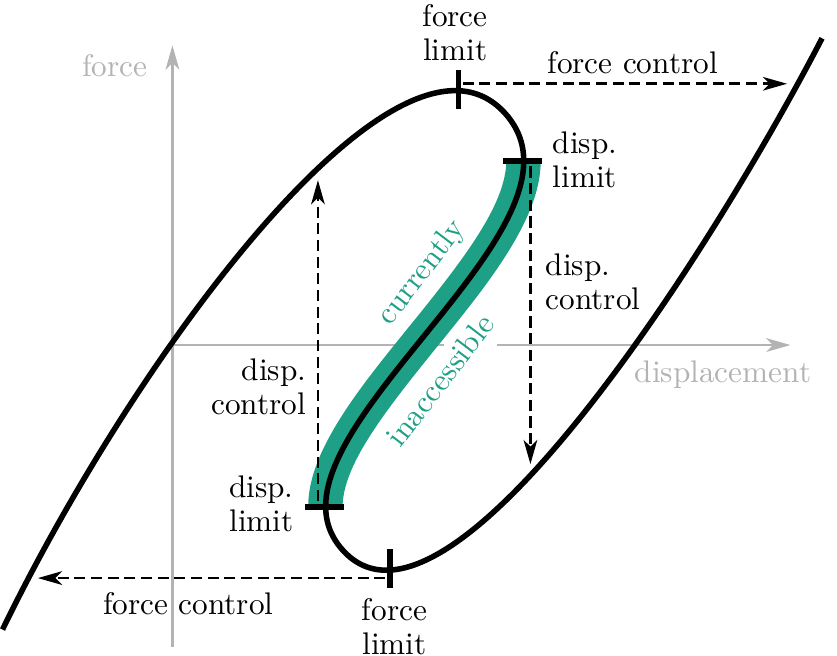}
    \caption{Limit points cause displacement and load control to snap to different parts of the equilibrium curve, resulting in an experimentally inaccessible region. 
    }
    \label{fig:snapback}
\end{figure}

Figure~\ref{fig:snapback} shows how force limit points cause force-controlled structures to snap to the next available stable equilibrium. Similarly, displacement limit points cause displacement-controlled structures to snap while conserving the displacement at the point(s) of actuation. To trace an equilibrium manifold like that shown in Figure~\ref{fig:snapback}, a means of controlling both the forces acting on the structure and its global deformation is required. This combination is readily implemented in a numerical setting because force and deformation can be independently controlled via a third parameter, namely the arc-length~\cite{Riks1979}. However, tracing similar equilibrium manifolds experimentally remains an open challenge. The challenge is that force and displacement at the actuation point(s) are not independent, but rather inherently linked through elasticity. A force applied at a specific control point results in a displacement at that point; similarly, an applied displacement induces a reaction force. This differentiates quasi-static from dynamic problems~\cite{Sieber2008} where the input vibration frequency and amplitude are decoupled.
   
In this paper, we present a general method to explore the unstable equilibria of a nonlinear structure quasi-statically. The key to this technique is decoupling force and displacement at a specific control point by introducing a third control variable: the shape of the structure. 

\begin{figure}
  	\centering
    \includegraphics{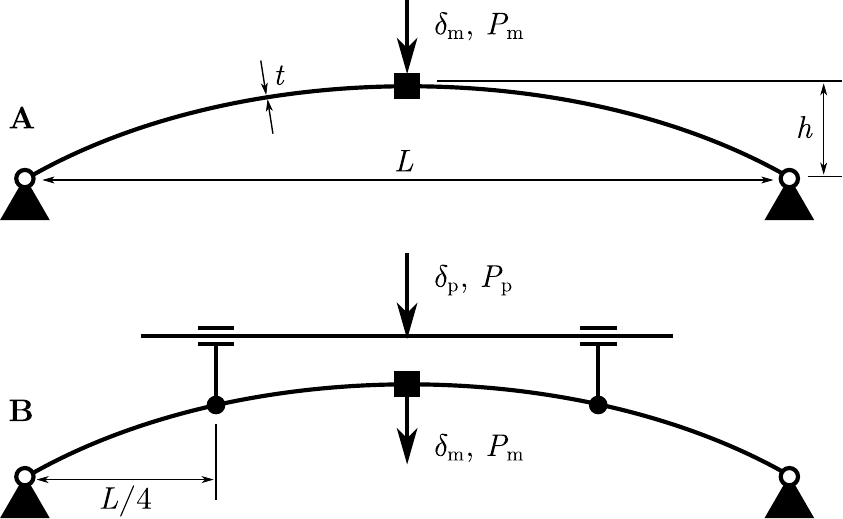}
    \caption{
    (A) The shallow arch structure studied in this paper. The edges of the arch are pinned. A vertical displacement ($\delta_\mathrm{m}$) or force ($P_\mathrm{m}$) is applied to the midpoint. Rotations and lateral translations are constrained at the midpoint to preserve symmetry.
    (B) Additional control points provide shape control. A displacement ($\delta_\mathrm{p}$) or force ($P_\mathrm{p}$) is applied symmetrically to the ``probes" halfway between the midpoint and edges. The probes allow rotations and lateral displacements in order to prevent reaction moments and horizontal reaction forces that would force the structure into a different equilibrium.
    }
    \label{fig:arch_config}
\end{figure}

\begin{figure*}
  	\centering
    \includegraphics{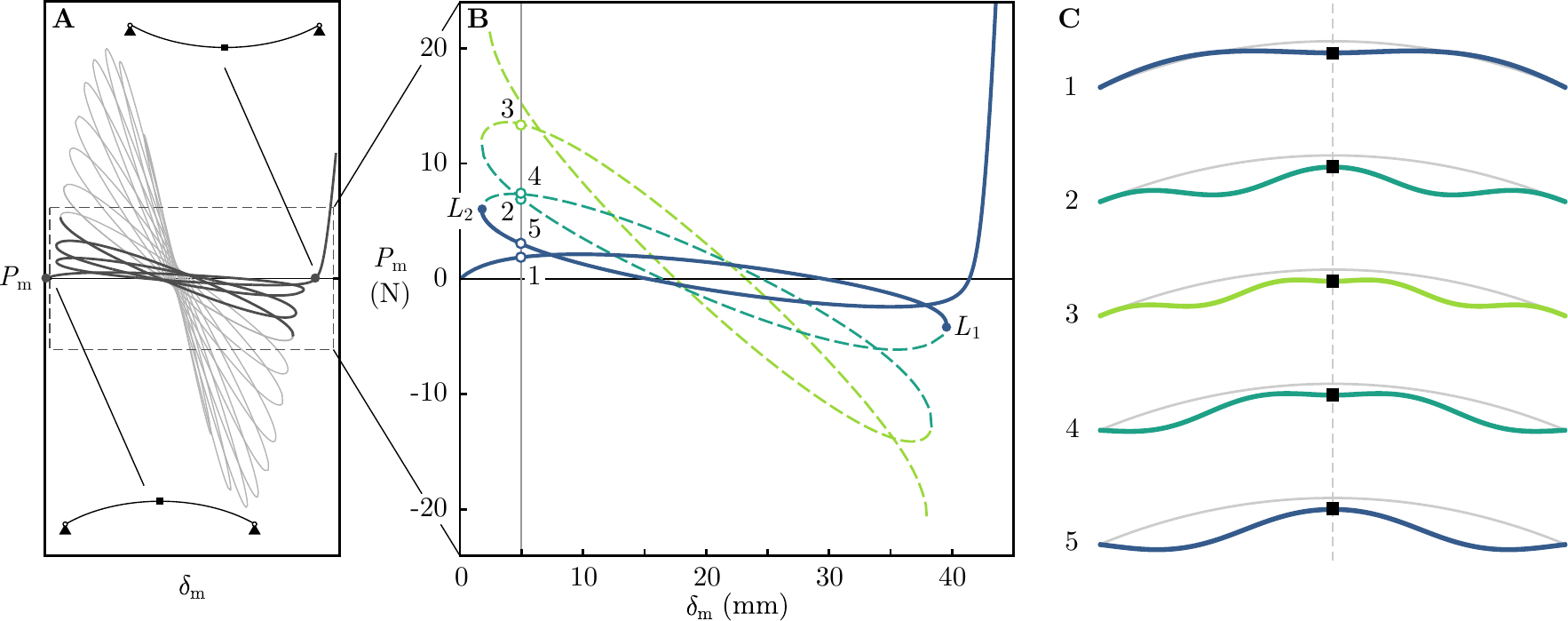}
    \caption{FE prediction of the highly nonlinear load-displacement behaviour of the shallow arch. 
    (A) The entirety of the symmetric response. The arch shapes corresponding to the first and final equilibria are shown. 
    (B) A subset of the response, with the solid blue lines indicating the segments a displacement-controlled experiment would obtain; dashed lines indicate equilibrium configuration currently inaccessible experimentally. At the limit points $L_1$ and $L_2$, displacement control snaps to the opposite blue segment of the equilibrium curve. At $\delta_\mathrm{m}=5$ there exist multiple load values for a given midpoint displacement (points 1--5). 
    (C) The arch shapes at points 1--5 for a fixed midpoint displacement.
    }
    \label{fig:FEA}
\end{figure*}

Consider a shallow arch loaded transversely at its midpoint as shown in Figure~\ref{fig:arch_config}A, which is a known benchmark for numerical arc-length solvers. For given combinations of geometry and material parameters, the arch features particularly pronounced nonlinear behaviour~\cite{Harrison1978} with many unstable loops in force-displacement space (Figure~\ref{fig:FEA}A). These loops give rise to the problem---accessing unstable equilibria and tracing experimentally inaccessible segments of equilibrium manifolds---but also the insight for a solution. For loops to exist, the structure must exhibit multiple different values of midpoint force for one midpoint displacement, and vice versa. By examination of the various arch shapes in Figure~\ref{fig:FEA}C and corresponding force values, it is evident that each force corresponds to a unique arch shape. Hence, each combination of midpoint force, midpoint displacement and deformation shape corresponds to a unique equilibrium. Control over the deformation shape is the key ingredient for decoupling force and displacement at the control point. For example, for a given midpoint displacement, the reaction force at the same point can be controlled \emph{indirectly} by changing the deformation shape.

The idea of separating force and displacement at a control point through shape control provides the first fundamental building block towards experimental continuation. The method for shape control is shown in Figure~\ref{fig:arch_config}B. Namely, we add two (to enforce symmetry) probe points, which allow us to manipulate the arch shape using another displacement-controlled input.

Experimental continuation requires feedback control interlinking loading and shape. Herein, we propose a simpler experiment as an initial step towards the full capability. Rather than moving the midpoint and probes simultaneously, we fix the midpoint at a given displacement and move the probes to scan for other equilibria. When the force on the probes reads zero, an equilibrium state of the system is found. With this method, which was recently used to determine localised solutions of the axially compressed cylinder~\cite{Thompson2017,Virot2017}, we find unstable equilibria which have never before been pinpointed quasi-statically. This concept of obtaining a zero-force reading on the probe to determine equilibria is analogous to the minimisation of virtual work in response to a probing virtual displacement, the vanishing of the residual in Newton's method at a converged state, and zero control in dynamic experimental continuation~\cite{Sieber2008}.

In previous work, the existence of unstable static equilibria has been intuited dynamically in the transients induced by large perturbations~\cite{WiebeVirgin2016}. Our approach differs in that the experimental setup stabilises otherwise unstable equilibria using additional control points. 
Additional control points have been used to constrain~\cite{Harvey2015} or probe~\cite{Virot2017} nonlinear structures in one direction, but are here rigidly fixed to the structure to allow the push/pull control required to scan for additional equilibria.

% =======================================================================

\section{Experimental Methods}
\label{sec:experimental}
\subsection{Experimental Setup and Equipment}
\begin{figure*}
  	\centering
    \includegraphics{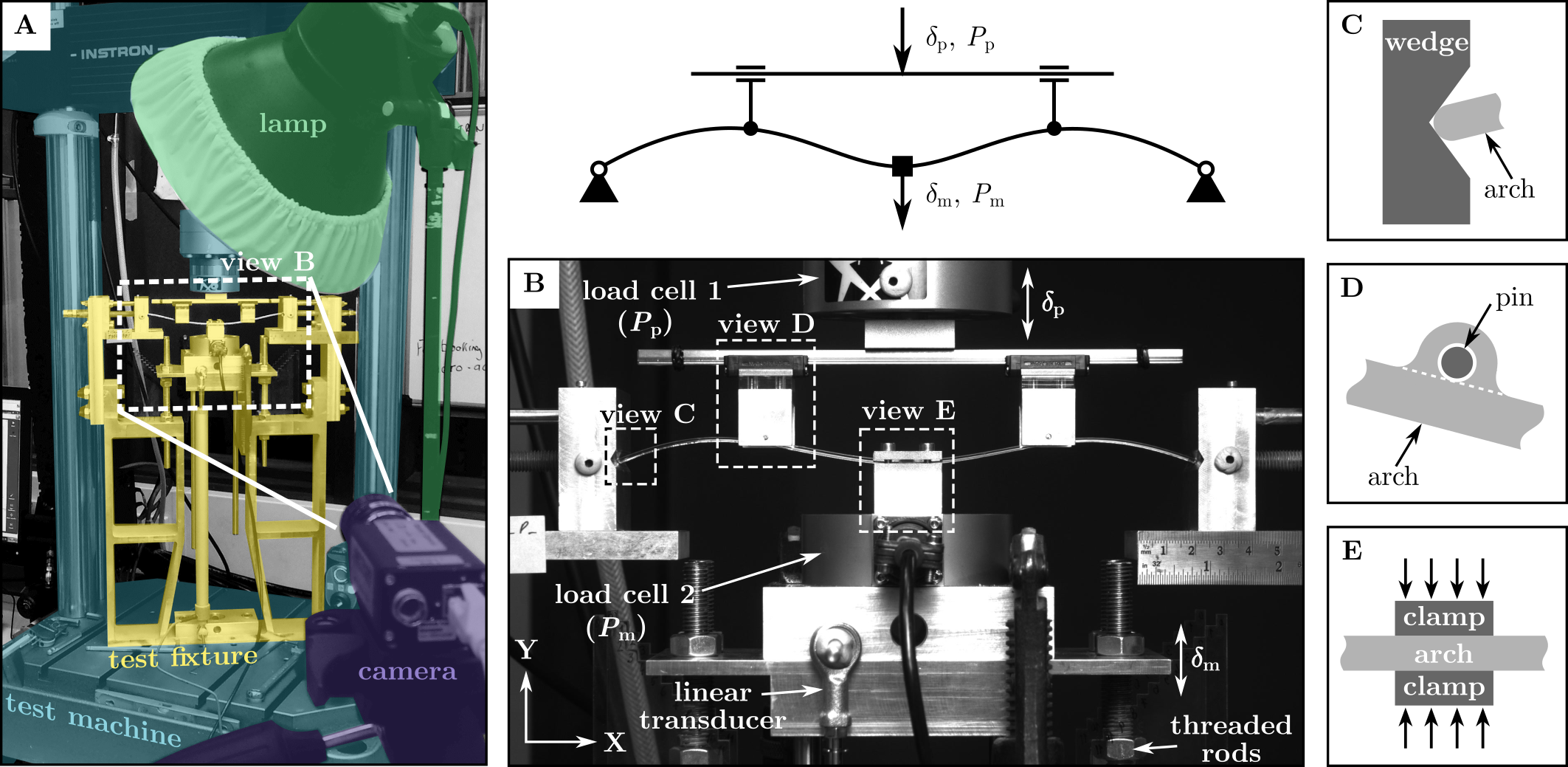}
    \caption{Experimental setup. 
    (A) The test fixture supporting the arch is bolted to the base of the test machine. The dashed rectangle shows the camera's field of view. 
    (B) Physical implementation of the arch loads and boundary conditions. The idealised model is shown above for reference. The test machine controls the probe displacement $\delta_\mathrm{p}$ and load cell 1 measures the probe force $P_\mathrm{p}$. 
    A moveable platform sets the midpoint displacement $\delta_\mathrm{m}$. A linear transducer measures the position of the midpoint platform. Load cell 2 measures the midpoint force $P_\mathrm{m}$. 
    (C) Wedge-shaped blocks restrain translations and allow rotations of the arch ends. 
    (D) A linear guide rail allows the probes to move in the X direction, while controlling the  displacement along Y. The probes connect to the arch with pins which allow rotations. 
    (E) A clamp restrains all translations and rotations at the arch midpoint to maintain symmetry. The clamped area is $5\,\mathrm{mm}$ wide. 
     }
    \label{fig:experimental_setup}
\end{figure*}

Figure~\ref{fig:experimental_setup} gives a detailed explanation of the experimental setup, including implementation of the loads and boundary conditions.

An Instron 8872 hydraulic test machine with an Instron Dynacell $\pm250\,\mathrm{N}$ tension/compression dynamic load cell (load cell 1) was used for all displacement-controlled tests. Load cell 2 was a $\pm500\,\mathrm{N}$ tension/compression load cell manufactured by Applied Measurements Ltd. A Gefran PZ34-A-250 linear transducer was used to measure the height of the midpoint platform. LabView (version 14.0) was used to log experimental data to ensure that synchronised readings were obtained from the various sensors. An Imetrum Video Gauge camera system was used to record the tests. 

\subsection{Arch Specimens}
\label{sec:arch_specimens}
The geometry of the arches tested is shown in Figure~\ref{fig:arch_config}, with dimensions $L=205\,\mathrm{mm}$, $h=20\,\mathrm{mm}$, $t=1.57\,\mathrm{mm}$, and depth $D=4.68\,\mathrm{mm}$ (into the page). Ten specimens were manufactured using a Trotec Speedy 100 laser engraver to cut the arches from sheets of acrylic (supplied by F.R.\ Warren Ltd). Mechanical coupon testing was performed to find the Young's modulus $E = 3200\pm70\,\mathrm{MPa}$ and Poisson's ratio $\nu = 0.38\pm0.02$ for use as inputs to Finite Element (FE) models. 

\subsection{Test Sequence}
Two types of tests were performed---\textit{midpoint scans} and \textit{probe scans}. Midpoint scans represent the standard displacement-controlled experimental approach. The test machine is connected to the midpoint clamp, and no probes are used. The midpoint is moved down and back up under displacement control. This produces a load-displacement curve similar to the solid blue lines in Figure~\ref{fig:FEA}B. The two segments of the equilibrium curve correspond to the ``downwards'' and ``upwards'' parts of the test. At limit points $L_1$ and $L_2$, the arch snaps to the other blue segment. A midpoint scan was performed for each of the $10$ arch specimens.

\begin{figure}
  	\centering
    \includegraphics{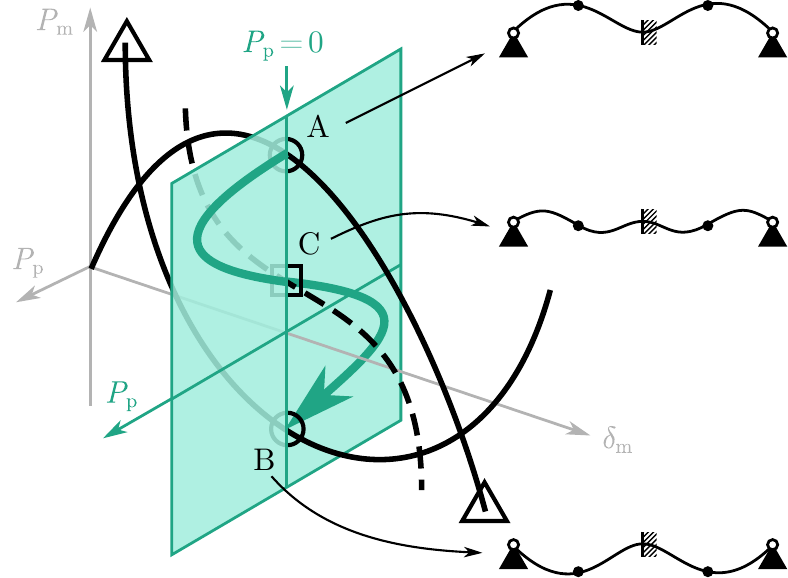}
    \caption{A schematic midpoint scan is plotted in black on the $P_\mathrm{m}$-$\delta_\mathrm{m}$ plane. The probe scan takes place on the green $P_\mathrm{p}$-$P_\mathrm{m}$ plane, which intersects the midpoint scan at a fixed value of $\delta_\mathrm{m}$. Starting at shape \textbf{A}, the probes move down to shape \textbf{B}. The probe reaction force, $P_\mathrm{p}$, is plotted on the green plane. For $P_\mathrm{p}=0$, an equilibrium configuration is detected. The probe scan detects the segments of the equilibrium curve which are stable with midpoint control only (\textbf{A} and \textbf{B}), and also detects an unstable segment (\textbf{C}).}
    \label{fig:probe-scan}
\end{figure}

In a probe scan, the configuration in Figure~\ref{fig:experimental_setup}B is used. The midpoint is fixed at a given displacement, and the probes are moved down and back up under displacement control. During this test, the arch passes through both stable equilibrium segments, and one or more unstable segments, as depicted in Figure~\ref{fig:probe-scan}. 

By repeating the probe scans at different midpoint displacements, the location of additional unstable equilibrium segments can be ``mapped out'' without path-following. For each arch specimen, a probe scan was performed for $\delta_\mathrm{m}= \{8,12,16,20,24,28,32\}\,\mathrm{mm}$. 

It is important to note that this method does not require us to follow or balance on an unstable equilibrium path; the structure is simply pushed \emph{through} an unstable equilibrium and its location on the midpoint load-displacement curve is measured. Consequently, the probe scan experiment can be performed by a displacement-controlled test machine. In future experiments, the probes will be controlled (via a more sophisticated feedback-control approach) to seek zero reaction force, and follow an unstable equilibrium segment while moving the midpoint. 

%=========================================================================

\section{Results}
\label{sec:results}

\subsection{Midpoint Scans}
\begin{figure*}
  	\centering
    \includegraphics{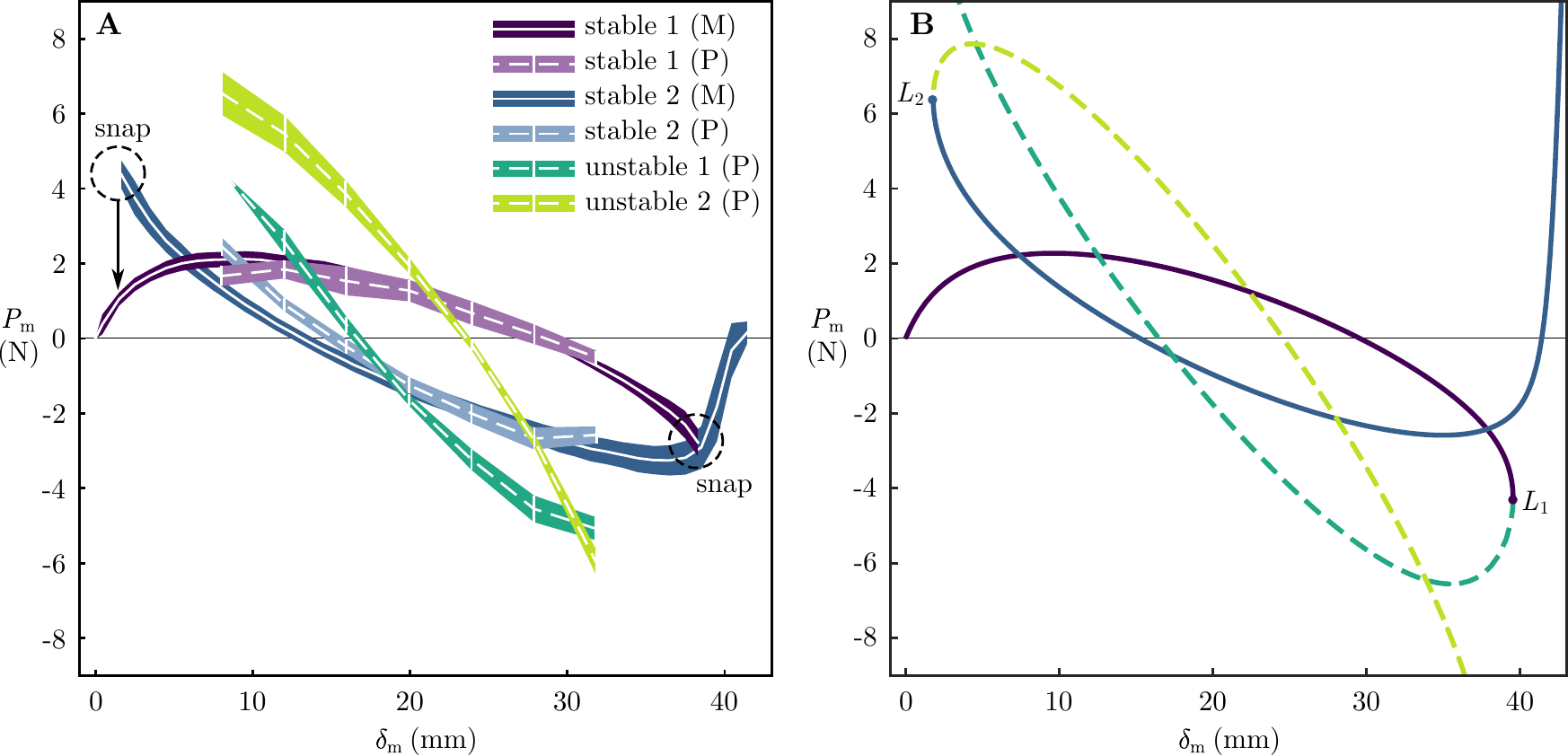}
    \caption{Comparison of experimental and FE results. 
    (A) Midpoint scan and probe scan results. Midpoint data after the snaps have been omitted. The white lines show the mean of the data, and the coloured areas represent $\pm$ one standard deviation of $P_\mathrm{m}$. The midpoint scans are indicated by an ``M'' in the legend; the probe scans  by a ``P''. The vertical white bars show the $\delta_\mathrm{m}$ locations of the probe scans. 
    (B) FE prediction of the midpoint symmetric load-displacement response, truncated to the ``first'' few segments as in Figure~\ref{fig:FEA}B. The segments are colour-coded to match their counterpart in the experimental results. }
    \label{fig:results}
\end{figure*}

The midpoint scan $\delta_\mathrm{m}$, $P_\mathrm{m}$ data from $10$ specimens were split into the ``downwards'' and ``upwards'' portions of the test, to prevent the loops in the plot affecting the following calculations. The data were then separated into 1~mm wide bins, and the mean and standard deviation of $\delta_\mathrm{m}$ and $P_\mathrm{m}$ were found for each bin. Figure~\ref{fig:results}A shows the results in purple and blue. The width of the filled area indicates the confidence interval of the measurements, based on the standard deviation.

\subsection{Probe Scans}
The probe scan $\delta_\mathrm{p}$, $P_\mathrm{p}$ data were analysed to find all the zero crossings of $P_\mathrm{p}$---i.e.\ the equilibria of the midpoint-controlled structure. The pairs of data points where $P_\mathrm{p}$ crossed zero were identified, and the $\delta_\mathrm{p}$ values corresponding to $P_\mathrm{p}=0$ were found by linear interpolation. The gradient of the load-displacement curve and the shape of the arch at the equilibrium point revealed the segment to which the equilibrium belonged. 

Two unstable segments were detected, in addition to the two known stable segments (Figure~\ref{fig:results}A). In all probe scans, one unstable equilibrium was found in the first (``downwards'') part of the test, and the other was found in the second (``upwards'') part of the test. Consequently, equilibria corresponding to the downwards/upwards part of the test are labelled with a  1 and a 2, respectively. 

At each $\delta_\mathrm{m}$ location, the equilibria detected by the probe scans were grouped by type, and the mean and standard deviation of $\delta_\mathrm{m}$ and $P_\mathrm{m}$ were calculated for each group. The aggregate data for each $\delta_\mathrm{m}$ location were then connected to give the mean $\pm$ standard deviation plot shown in Figure~\ref{fig:results}A. The results of the probe scans match the results of the midpoint scans fairly well, despite the fact that the two types of tests are performed separately, and with different equipment set-ups. Consequently, it is assumed that the unstable segments have been located with a similar degree of accuracy. 

Discrepancies between the probe and midpoint scan results may be due to relaxation of the specimen material. Upon removing the arch from the fixture after the final probe scan, it was noted that most specimens did not immediately spring back to their undeformed shape. Imperfections in the experimental set-up (especially the boundary conditions) may also have contributed to the differences.

An FE analysis of the arch was performed using nonlinear beam elements and idealised pinned boundary conditions.
Figure~\ref{fig:results}B shows the FE prediction, with the segments colour-coded to match their experimental counterparts in Figure~\ref{fig:results}A. This shows that we have located the ``next'' two segments of the arch response---i.e.\ the segments beyond limit points $L_1$ and $L_2$. These segments correspond to arch shapes which are not stable under midpoint control only (shapes 2 and 4 in Figure~\ref{fig:FEA}C), but are stable when supported by the probes. Further segments will correspond to more complex shapes (e.g.\ shape 3 in Figure~\ref{fig:FEA}C), which will require additional probes for support. 

Despite the sensitivity to initial conditions in nonlinear systems, there is excellent qualitative and quantitative agreement between the experimental and theoretical results. This provides confidence that the unstable equilibria have been correctly identified by the testing method.

% ==================================================================

\section{Conclusions and Outlook}
\label{sec:conclusion}
We have presented an experimental method to detect and identify unstable equilibria of nonlinear structures quasi-statically, using probes to control the shape of the structure. Using this method, we have, for the first time experimentally, shown the location of unstable equilibria of a shallow arch which would not be accessible using traditional quasi-static testing techniques. These equilibria correspond to structural shapes that have zero reaction force at the probe points. The shape control provides independent, albeit indirect, control over force and displacement at the point of actuation, which are otherwise intrinsically linked.  

The natural extension of this work is to exploit the probing technique as a means to trace equilibrium paths of nonlinear structures. This may be achieved through a concerted variation in actuation point force or displacement, and shape control via the probes. The result is an experimental continuation technique, which enables the quasi-static nonlinear response of a structure to be explored systematically. In future, the addition of multiple, independent probes would provide more refined control over the structural shape, thereby enabling more segments of the equilibrium manifold to be identified experimentally.

The development of continuation techniques using shape control will enable the experimental validation of the response of nonlinear structures. In turn, this will help encourage the exploitation of nonlinear structures in engineering applications, for example in morphing structures and compliant mechanisms.

\section{Acknowledgments} 
This work was funded by the EPSRC through grant numbers EP/N509619/1, EP/M013170/1 and EP/P511298/1. EPSRC's support is gratefully acknowledged. The authors also wish to acknowledge the help of the University of Bristol technicians: G.~Pearn, A.~Kraft, R.~Billingham, R.~Bragg, and R.~Hooper. 

\bibliography{experimentalcontinuation}

\end{document}